\documentclass[12pt]{article}
\usepackage{setspace}
\usepackage{amsmath,amssymb}
\usepackage{color}
\usepackage{hyperref}

\begin{document}
\setlength{\topmargin}{-1cm} 
\setlength{\oddsidemargin}{-0.25cm}
\setlength{\evensidemargin}{0cm}

\newcommand{\e}{\epsilon}
\newcommand{\beq}{\begin{equation}}
\newcommand{\eeq}[1]{\label{#1}\end{equation}}
\newcommand{\bea}{\begin{eqnarray}}
\newcommand{\eea}[1]{\label{#1}\end{eqnarray}}
\renewcommand{\Im}{{\rm Im}\,}
\renewcommand{\Re}{{\rm Re}\,}
\newcommand{\diag}{{\rm diag} \, }
\newcommand{\Tr}{{\rm Tr}\,}
\def\draftnote#1{{\color{red} #1}}
\def\bldraft#1{{\color{blue} #1}}

\begin{titlepage}
\begin{center}

\vskip 4 cm

{\Large \bf  Notes on Relevant, Irrelevant, Marginal and Extremal Double Trace Perturbations }

\vskip 1 cm

{Massimo Porrati \footnote{E-mail: \href{mailto:mp9@nyu.edu}{mp9@nyu.edu}}  and Cedric C.Y. Yu \footnote{E-mail: \href{mailto:cyy272@nyu.edu}{cyy272@nyu.edu}}}

\vskip .75 cm

{\em Center for Cosmology and Particle Physics, \\ Department of Physics, New York University, \\ 4 Washington Place, New York, NY 10003, USA}

\end{center}

\vskip 1.25 cm
%\title{}
%\author{}
%\date{}Killing
%\maketitle

\begin{abstract}
\noindent  Double trace deformations, that is products of two local operators, define perturbations of conformal field theories
 that can be studied exactly in the large-$N$ limit. Even when the double trace deformation is irrelevant in the infrared, it is 
 believed to flow to 
 an ultraviolet fixed point.  In this note we define the K\"allen-Lehmann representation of the  two-point function of a local  
 operator $O$ in a theory perturbed by the square of such operator. We
use such representation to discover potential
 pathologies at intermediate points in the flow that may prevent to reach the UV fixed point. We apply the method to
 {an ``extremal''} deformation that naively would flow to a UV fixed point where the operator $O$ would saturate the unitarity bound   
 $\Delta=\frac{d}{2}-1$. We find that the UV fixed point is not conformal and that the deformed two-point function propagates unphysical modes. We interpret the
 result as showing that the flow to the UV fixed point does not exist. This resolves a potential puzzle in the holographic
 interpretation of the deformation. 
\end{abstract}
\end{titlepage}
\newpage

\section{Introduction}
Certain conformal field theories that come in families defined by an integer $N$ simplify in the large-$N$ limit. Notable 
examples are theories that can be described by weakly coupled gravities via the holographic duality and $O(N)$ vector 
models. The latter are deformations of $N$ massless free field theories. The deformation is called by abuse of term
 ``double trace.'' It is defined as $\lambda N O^2$, with  $O= C\sum_{a=1}^N \phi^a\phi^a$ (for bosons) or
${{}}{O}=C\sum_{a=1}^N \bar{\psi}^a\psi^a$ (for fermions). Choosing $C={\cal O}(1/N)$ the correct large-$N$ 
scaling for the coupling $\lambda$ of the double-trace perturbation is $\lambda={\cal O}(N^0)$. 

The deformed
$O(N)$ models are particularly interesting in three spacetime dimensions. The deformation of the bosonic model is 
relevant and flows to a nontrivial infrared fixed point~\cite{Brezin:1972se,Wilson:1973jj}. The deformation of the fermionic model
is irrelevant, so non-renormalizable by power counting. Nevertheless, a UV fixed point is believed to exist in the
large-$N$ limit~\cite{Gross:1974jv,Parisi:1975im,rw} and even at finite 
$N$~\footnote{See~\cite{Fei:2016sgs} for a recent review of evidence in favor of the existence of a UV fixed point.}. Both models can be solved exactly by using a method that easily generalizes to the case of
adjoint theories with holographic semiclassical gravity duals. In the context of holographic duality, the correct treatment of
multi-trace perturbations was explained in~\cite{Witten:2001ua} and further simplified in~\cite{Mueck:2002gm}. The 
analysis of~\cite{Mueck:2002gm} was further extended and generalized beyond AdS/CFT holography 
in~\cite{Elitzur:2005kz}. We will review the results of~\cite{Mueck:2002gm,Elitzur:2005kz} --for completeness 
and to fix notations and normalizations-- in section 2. 
 
 The rediscovery of multi-trace perturbations in the context of AdS/CFT duality makes clear that they can be studied exactly
 in an appropriately defined large-$N$ limit~\footnote{ Here $N$ denotes the order of magnitude of the number of degrees of freedom in the unperturbed CFT. In even dimensions it is given, at least parametrically, by the coefficient of a conformal anomaly.}, 
even when the CFT is not free and the operator $O$ has arbitrary conformal
 dimension $\Delta$. A feature of the exact solution of the deformed model is that, at the fixed points, 
 the conformal dimension of a deformation which is the product of primary fields $O_1,... O_m$ is the sum of the
 individual dimensions $\Delta=\Delta_1+...+\Delta_m$, but the dimensions of the individual operators at the two fixed points (UV and IR) are in general 
different. 
 
 The possibility of solving the deformed model raises several 
 interesting questions, which we shall try to answer in this paper.
One such question arises already in the case of a double-trace deformation $O^2$ when $2\Delta>d$ in $d$ spacetime 
dimensions. In this case the deformation is irrelevant in the IR, that is non renormalizable by power counting. Nevertheless
the exact solution of the deformed model at large ``$N$" 
has a UV fixed point. A difficult question is whether this UV fixed 
point exists at finite $N$. A simpler one is whether the fixed point can be connected to the IR by a physical renormalization
group (RG) flow. By ``physical'' we mean the following: the deformed theory has both an IR and a UV fixed 
point; therefore, it defines a field theory valid at all energies. The theory is thus a UV complete one rather than merely an 
effective field theory, valid only up to a maximum energy scale. To qualify as a ``physical'' this theory must be free of 
tachyons and ghosts. Of course an ``unphysical'' theory, plagued by ghosts or tachyons, can still have physical UV or
IR fixed points. We will see several examples of such behavior in section 3. Such theory may also describe an interesting statistical mechanics system, but not a relativistic, local field theory. We call an RG flow ``physical'' when it is generated
by a deformation that produces a physical relativistic field theory. Ref.~\cite{Witten:2001ua} shows that the RG flow due to the double trace perturbation $O^2$  connects
a fixed point where the scaling dimension of $O$ is $\Delta_+=d/2+\nu$ to one where the dimension is
$\Delta_-=d/2-\nu$. In the range $0<\nu<1$ both dimensions satisfy the unitarity bound~$\Delta\geq d/2-1$. Outside
the range, $\Delta_-$ violates the unitarity bound, so the RG flow should be pathological. This turns out to be the case: the 
pathology shows up in a particular UV completion of the theory 
because of the presence of unphysical poles in the two-point function of 
the operator 
$O$~\cite{Andrade:2011dg}. 

We will recover this pathology in section 3, where we will refine the analysis of 
refs.~\cite{Andrade:2011dg,Andrade:2011aa} by giving a complete K\"allen-Lehmann representation of the two-point
correlator $\langle O(x) O(0)\rangle$. Section 3 will also study the case $\nu=1$, which is perhaps the most intriguing of all
 double-trace perturbations for CFTs that are dual to semiclassical gravity theories. Their bulk dual
 contains a scalar, $\Phi$, with squared mass $(mL)^2= -d^2/2 +1$ ($L$=AdS radius). 
 The standard quantization of such bulk theory associates to the bulk scalar $\Phi$ an operator $O$ of dimension
 $d/2+1$. The RG flow, if it existed, would end in a theory where $O$ had dimension $d/2-1$, which
therefore would saturate the unitarity bound. In a unitary 
 CFT any operator saturating the unitarity bound must be a free field~\cite{Buchholz:1976hz}~\footnote{See 
 also~\cite{Weinberg:2012cd}, and~\cite{Simmons-Duffin:2016gjk} for a recent review.}. This free field should come 
 from a different identification of sources and VEVs in a theory 
with the same near-boundary behavior of the scalar field and the same bulk 
 action~\cite{Klebanov:1999tb}. On the other hand, a standard scalar action does not carry the singleton representation 
 corresponding to a free 
scalar~\cite{Breitenlohner:1982jf}~\footnote{A dipole
action for the singleton on a {\em fixed} $AdS_4$ background was proposed 
in~\cite{FF}. The action is quadratic and propagates no degrees of freedom in 
the bulk.}. So, an obvious question to ask is whether the flow induced 
 by the {``extremal'' double trace perturbation $O^2$, with $O$ of dimension $d/2+1$,} is physical all the way up to 
 the UV and really terminates in a  free-field fixed point. We will be able to
 answer this question (in the negative) at the end of section 3. Final remarks on operator mixing and connections to other papers on double-trace perturbations --in section 4-- and an appendix summarizing 
 the AdS/CFT holographic description of multi-trace perturbations conclude the paper. 
 
 \section{Multi Trace Deformations}

As a warm-up example  of multi-trace deformations let us consider the interacting $O(N)$ vector model. 
In three dimensions it was conjectured to be the holographic dual of $AdS_4$ high spin theories 
in~\cite{Klebanov:2002ja}. Its action is

\beq
I_{CFT}=\int d^dx \sum_{a=1}^N \left(\frac{1}{2}(\partial \phi^a)^2+
\frac{\lambda}{2N}(\phi^a \phi^a)^2 \right),\;\;\lambda>0.
\eeq{m1}
The deformation is relevant in dimension $d<4$.

This action can be expressed in terms of the bilinear $\phi^a \phi^a$ by introducing an auxiliary field $\Sigma$:
\beq
I_{CFT}=\int d^dx \sum_{a=1}^N \left(\frac{1}{2}(\partial \phi^a)^2+ \lambda(\phi^a \phi^a)\Sigma-{1\over 2}N\lambda \Sigma^2 \right)
\eeq{m2}
in which the expectation value $\langle \Sigma\rangle=\langle \sum_{a=1}^N N^{-1} \phi^a \phi^a\rangle$ is
formally ${\cal O}(N^0)$, i.e.  finite in the large-$N$ limit. Integrating out 
$\Sigma$ one recovers action~(\ref{m1}); integrating
out $\phi^a$ {and discarding a $\Sigma$-independent constant,} one obtains instead a non-local action for $\Sigma$  
\beq
S[\Sigma]={N\over 2} \log\det [-\partial^2/2 +\lambda\Sigma] -
{N\lambda \over 2}\int d^d x \Sigma^2 \equiv N s[\Sigma].
\eeq{m3} 
The ``intensive'' action $s[\Sigma]$ is independent of $N$. 
The generator of connected correlators of the operator $\sum_{a=1}^N  \phi^a\phi^a$, $W[J]\equiv Nw[J]$ is defined by

\beq
Z[J]=e^{N w[J]}=\int [d\Sigma]  \exp\left[ -{N\over 2}  \log\det [-\partial^2/2 + 
\lambda\Sigma -J] +{N\lambda \over 2}\int d^d x \Sigma^2 \right].
\eeq{m4} 
It is convenient to define an effective action $\gamma_\lambda[O]$, independent 
of $N$ up to terms ${\cal O}(1/N)$, as the Legendre transform 
of $w[J]$:
\beq
\gamma_\lambda[O]=\int d^dx OJ -w[J] , \qquad O(x)={\delta w\over \delta J(x)}.
\eeq{m4a}
In the large-$N$ limit the  integration over $\Sigma$ in 
eq.~(\ref{m4}) reduces to computing a saddle point. Therefore, the effective
action is 
\beq
\gamma_\lambda[O]= \int d^dx JO +{1\over 2} \log\det [-\partial^2/2 + \lambda\Sigma] 
-{\lambda \over 2}\int d^d x (\Sigma+J/\lambda)^2 ,
\eeq{m3a}
computed at the stationary point in $J$ and $\Sigma$. By computing first the
stationary point in $J$, it is easy to find that 
effective action of the deformed $O(N)$ model is
\beq
\gamma_\lambda[O]=\gamma_0[O] + {\lambda\over 2}\int d^dx O^2.  
\eeq{m4aa}
So, the effect of the deformation is additive in the effective action. Notice
that this result follows simply from the fact that at large $N$ the integral in
$\Sigma$ can be evaluated using the saddle point approximation.
 
As we mentioned before, $W[J]=Nw[J]$ generates connected correlators of the operator $\sum_{a=1}^N \phi^a \phi^a$, which are 
all ${\cal O}(N)$. This
is another manner of checking that $w[J]$ is independent of $N$ in the 
large-$N$ limit. Notice that the field $O$ appearing in the free energy is the 
expectation value
 of the normalized operator $\sum_{a=1}^N N^{-1} \phi^a\phi^a$, 
which differs from the operator sourced by $J$ by the normalization factor 
$N^{-1}$.   

We see that the double-trace perturbation is additive in the effective action 
at leading order in $1/N$. This simple result generalizes easily 
to any multi-trace deformation 
$NU(\sum_{a=1}^N N^{-1}\phi^a \phi^a)$, when the function
$U(x)$ is independent of $N$~\cite{Amit:1984ri}, and  to any theory admitting 
a large-$N$ limit~\cite{Elitzur:2005kz}. 

In fact, in all theories with an effective action 
${\cal O}(N^*)$, with $O$ normalized to be ${\cal O}(1)$, 
the  perturbation $\int d^dx N^*U(O)$ shifts the effective action
from the unperturbed value $\Gamma=N^*\gamma(O)$ to 
$\Gamma_U(O)\equiv N^*\gamma(O) + \int d^dx N^*U(O)$. 
Here $N^*$ is a (large) number counting the effective degrees 
of freedom of the theory. In even dimensions, this is proportional to the 
coefficient one of the the conformal anomalies. For $O(N)$ vector models 
$N^*=N$ while for CFTs with fields in the adjoint representation of a rank-$N$
algebra, such as those that possess holographic duals $N^*={\cal O}(N^2)$. 

To prove additivity of the multi-trace perturbation we begin by writing the Feynman integral 
representation of the free energy in Lorentzian signature, using 
the functional Fourier transform of the Dirac delta function.
We will denote by $\phi$ the fundamental fields of the CFT and we will use 
the notation $<A,B>\equiv\int d^dx A(x)B(x)$ henceforward. 
\beq
\exp(-iW[J])=\int [d\phi dt d\Omega] \exp\left[ +iI[\phi] -i<N^*,U(\Omega)> 
-iN^*<J,\Omega> + 
i<t,\Omega-O[\phi]>\right].    
\eeq{mm1}
The composite operator $O[\phi]$ is normalized so that 
$\langle O[\phi]\rangle={\cal O}(1)$ for $N^*\gg 1$.  
The functional Fourier transform 
\beq
\int [dJ]\exp(-iW[J] + i N^*<J,O>)\equiv \exp(i\Gamma_U[O]),
\eeq{mm2} 
defines the functional $\Gamma_U[O]$ in the perturbed theory in terms of the 
unperturbed functional $\Gamma[O]$ as
\beq
 \exp(i\Gamma_U[O])= \int [d\phi] \exp\left[ iI[\phi] -i<N^*,U(O)>\right] 
\delta[O-O[\phi]]=\exp(i\Gamma[O]-i<N^*,U(O)>).
\eeq{mm3}
So far all manipulations have been formal, but exact in $N^*$. In the large $N^*$ limit, any theory in which the free energy
is $W[J]=N^*w[J] + {\cal O}(1)$, with $w[J]$ independent of $N^*$, possesses two additional properties: 1)
$\Gamma[O]= N^*\gamma[O] + {\cal O}(1)$, with $\gamma[O]$ independent of $N^*$; 2) $\gamma[O]$ is the Legendre transform of $w[J]$.
The first property is obvious and the second follows from the saddle-point approximation of the functional integral~(\ref{mm2}). The second property also
identifies $\Gamma[O]_U$  with the effective action of the
perturbed theory and $\Gamma[O]$ with the effective action of the unperturbed
theory. 
\section{Two-Point Functions of Double Trace Perturbations and their K\"allen-Lehmann Representation}
Consider an operator $O$ of general conformal dimension $\Delta=d/2+\nu$. Unitarity 
requires $\Delta\geq d/2-1$ (i.e. $\nu\geq -1$); when the bound is saturated, $\Delta=d/2-1$, ${O}$ is necessarily 
free~\cite{Buchholz:1976hz}  (see also~\cite{Weinberg:2012cd,Simmons-Duffin:2016gjk}). 

Without any deformation, the connected two-point function of ${O}$ is 
\beq
\langle {O}(x){O}(0) \rangle = \frac{K}{|x|^{2(d/2+\nu)}} \equiv K G_\nu(|x|).
\eeq{m5}
The positive pre-factor $K$ is ${\cal O}(1/N^*)$, when the operator $O$ is normalized as in the previous section. 
In momentum space, defining $\widetilde{O}(k)=\int d^dx e^{-ik\cdot x}{O}(x)$, the two-point function is~\cite{Klebanov:1999tb}
\beq
\langle \widetilde{O}(k){O}(0) \rangle=K \widetilde{G}_\nu(k^2)=K C\left({k^2\over 4}\right)^\nu
\eeq{m6}
for non-integer $\nu$. The coefficient $C=\pi^{d/2} \Gamma(-\nu)/\Gamma(\nu+d/2)$ is negative for $\nu\in(2m,2m+1)$ (in particular $1>\nu>0$) and 
positive for $0>\nu\geq -1$ and $\nu\in (2m+1,2m+2)$, $m\in \mathbb{Z}$.

The cases where $\nu \in \mathbb{Z}^+$ need to be considered separately 
due to the appearance of $\ln{k^2}$ terms. In particular, 
for $\Delta=d/2+1$ one finds (see Appendix)
\beq
\widetilde{G}_{\nu=1}(k^2)=K^{-1} \langle \widetilde{{{}}{O}}(k){{O}}(0) \rangle=C' k^2 \ln{(k^2/\mu^2)}
\eeq{m7}
with $C'$ positive for $d>2$ and $\mu$ an arbitrary scale that can be changed by adding a contact term proportional to $k^2$. 

Similarly, for $\Delta=d/2$, 
\beq
\widetilde{G}_{\nu=0}(k^2)=-C''\ln{(k^2/\mu^2)}\;,\;C''>0.
\eeq{m8}

The undeformed effective action is then
\beq
\gamma_0[O]= \frac{1}{2}\int \frac{d^dk}{(2\pi)^d} \widetilde{O}(k) \frac{1}{\widetilde{G}_{\nu}(k^2)} \widetilde{O}(-k).
\eeq{m9}

Now we add a double-trace deformation to the effective action $\gamma[O]$, 
\beq
U[O]=+\frac{\lambda}{2\Lambda^{2\nu}} O^2
\eeq{m10}
where $\lambda$ is dimensionless. The cases $\lambda>0$ and $\lambda<0$ will be considered separately. This deformation is IR-relevant for $\Delta<d/2$, marginal for $\Delta=d/2$ and irrelevant for $\Delta>d/2$. It is tempting 
to identify $\Lambda$ with the cut-off of the theory;  the rest of 
this section will substantiate such identification.

The deformed effective action becomes
\beq
\gamma[O]=+\frac{1}{2}\int \frac{d^dk}{(2\pi)^d}\widetilde{O}(k)\left(\frac{1}{\widetilde{G}_\nu(k^2)}+\frac{\lambda}{\Lambda^{2\nu}}\right)\widetilde{O}(-k).
\eeq{m11}

Thus we obtain the deformed two-point function 
\beq
\widetilde{G}_{\nu,\lambda}(k^2)\equiv \frac{1}{\frac{1}{\widetilde{G}_\nu(k^2)}+\frac{\lambda}{\Lambda^{2\nu}}}.
\eeq{m12}

\subsection{Case 1: $1>\nu>0$}
It was pointed out in \cite{Witten:2001ua} that for $1>\nu>0$ such a double-trace deformation leads to an RG flow in which the IR and UV fixed points are CFTs 
in which the operator $O$ has conformal dimension $\Delta=\Delta_\pm=d/2\pm \nu$ respectively. They are the two possible choices of quantization in the AdS/CFT context for a massive scalar in the bulk~\cite{Klebanov:1999tb} (see Appendix for a review). This flow can be achieved with $\lambda$ positive or negative. 
Recall that, 
omitting a positive coefficient, $\widetilde{G}_\nu(k^2)=- k^{2\nu}$. 
In the IR regime, $k^{2\nu}\ll \Lambda^{2\nu}/|\lambda|$, 
\bea
\widetilde{G}_{\nu,\lambda}(k^2)&=&-k^{2\nu}\frac{1}{1-\frac{\lambda  k^{2\nu}}{\Lambda^{2\nu}}}\nonumber\\
&\approx& -k^{2\nu} \left( 1+\frac{\lambda  k^{2\nu}}{\Lambda^{2\nu}} \right)\nonumber\\
&\approx& -k^{2\nu}\\
&=& \widetilde{G}_{\nu,\lambda=0}(k^2)\nonumber
\eea{m13}
which reduces to the original undeformed CFT of $\Delta=\Delta_+$, as expected of an irrelevant deformation.

In the UV, on the other hand, $k^{2\nu}\gg \Lambda^{2\nu}/|\lambda|$, 
\bea
\widetilde{G}_{\nu,\lambda}(k^2)&=&\frac{\Lambda^{2\nu}}{\lambda}\frac{1}{1-\frac{\Lambda^{2\nu}}{\lambda k^{2\nu}}}\nonumber\\
&\approx& \frac{\Lambda^{2\nu}}{\lambda} \left(1+\frac{\Lambda^{2\nu}}{\lambda k^{2\nu}} \right)\nonumber\\
&=& \left(\frac{\Lambda^{2\nu}}{\lambda}\right)^2 \frac{1}{k^{2\nu}}+\mbox{contact terms}\\
&\propto& \widetilde{G}_{-\nu,\lambda=0}(k^2).\nonumber
\eea{m14}
Upon removing the contact term, this is the two-point function of a CFT  with $\Delta=\Delta_-$.

Now we come to the heart of our paper. We 
express the two-point function in K\"allen-Lehmann form. 

\subsubsection{$\lambda<0$}

Consider first $\lambda<0$. \color{black}

\bea
\widetilde{G}_{\Delta,\lambda<0}(k^2)&=&-\frac{1}{\frac{1}{k^{2\nu}} +  \frac{|\lambda|}{\Lambda^{2\nu}}}\nonumber\\
&=&-\frac{\Lambda^{2\nu}}{|\lambda|}\left( 1-\frac{1}{|\lambda| (k^2/\Lambda^2)^{\nu} +1} \right)\nonumber\\
&=& +\frac{\Lambda^{2\nu}}{|\lambda|}\left( \frac{1}{|\lambda| (k^2/\Lambda^2)^{\nu} +1} \right)+\mbox{contact terms}.
\eea{m15}
Let us introduce now the complex function $f(z)=\frac{1}{|\lambda| z^\nu+1}$; 
when its branch cut is placed on the negative real axis 
it is meromorphic in the range ($\pi\geq\arg{z}>-\pi$).

In this case $f(z)$  has no singularity in the first sheet. 
This implies that there are no tachyonic or otherwise unphysical one-particle
states
among the states created by applying ${O}(x)$ to the vacuum. 
Using Cauchy's formula, one finds
\beq
\widetilde{G}_{\nu,\lambda<0}(k^2)=+\frac{\Lambda^{2\nu}}{\pi}\int_0^\infty \frac{dm^2}{k^2+m^2}\frac{(m^2/\Lambda^2)^\nu \sin{\pi \nu}}{1+\lambda^2 (m^2/\Lambda^2)^{2\nu}+2|\lambda| (m^2/\Lambda^2)^\nu \cos{\pi\nu}}.
\eeq{m16}

So,
the spectral density is positive-finite and the spectrum is free of ghosts and tachyons. Thus the deformation with $\lambda<0$ may
provide a healthy flow between the IR and UV fixed points. Of course there is
a (nonperturbative) fly in the ointment here, since $\lambda<0$ means that the
potential $\lambda O^2$ is unbounded from below.

\subsubsection{$\lambda>0$}

Consider next the case where $\lambda>0$:

\beq
\widetilde{G}_{\nu,\lambda>0}(k^2)= \frac{\Lambda^{2\nu}}{\lambda}\left( \frac{1}{\lambda (k^2/\Lambda^2)^{\nu} -1} \right),
\eeq{m18}
which has a pole of positive residue at $k^2/\Lambda^2=\lambda^{-1/\nu}$ on the positive real axis, signaling a (non-ghost) tachyon mode. The K\"allen-Lehmann 
representation shows that the continuum part of the spectral density is
positive definite, so the tachyon is the only unphysical feature of the 
deformed theory: 
\bea
\widetilde{G}_{\Delta,\lambda>0}(k^2)&=&\frac{\Lambda^{2\nu}}{\lambda}\frac{1}{\nu \lambda^{1/\nu}\;(k^2/\Lambda^2-\lambda^{-1/\nu})}\nonumber\\
&\quad&+\frac{1}{\pi}\int_0^\infty \frac{dm^2}{k^2+m^2}\frac{m^{2\nu} \sin{\pi \nu}}{1+\lambda^2 (m^2/\Lambda^2)^{2\nu}-2\lambda (m^2/\Lambda^2)^\nu \cos{\pi\nu}}.
\eea{m19}

Two limits are worth mentioning. The first is $\Lambda\rightarrow \infty$. In
this case the tachyon moves to infinite mass and the perturbation disappears
(since it becomes irrelevant at all energy scales).

The second limit is less trivial. It is the UV limit in which 
$\Lambda\rightarrow 0$. One interesting question is whether the UV limit
may exist as a CFT even if the deformation leading to it is unphysical.
The answer is yes, because in that limit  
\bea
\widetilde{G}_{\nu,\lambda>0}(k^2)&\approx& \frac{\Lambda^{2\nu+2}}{\lambda}\frac{1}{\nu \lambda^{1/\nu}\;k^2}+\frac{1}{\pi}\int_0^\infty \frac{dm^2}{k^2+m^2}\frac{\Lambda^{4\nu}m^{2\nu} \sin{\pi \nu}}{\lambda^2 m^{4\nu}}\nonumber\\
&{\approx}&\frac{1}{\pi}\int_0^\infty \frac{dm^2}{k^2+m^2}\frac{\Lambda^{4\nu}\sin{\pi \nu}}{\lambda^2 m^{2\nu}}.
\eea{m20}
In other words, the tachyonic mode decouples.

\subsection{Case 2: $0>\nu>-1$}
One natural question to ask is: Do we get a flow similar to the one above by adding a double-trace deformation for the operator associated to the alternative quantization $d/2>\Delta>d/2-1$ ($0>\nu>-1$)?

In this case 
\beq
U[O]=+\frac{\lambda}{2\Lambda^{2\nu}} O^2
\eeq{m52}
is IR-relevant (and UV-irrelevant).

\subsubsection{$\lambda>0$}

For $\lambda>0$, 
\beq
\widetilde{G}_{\nu,\lambda>0}(k^2)= \frac{1}{\frac{1}{k^{2\nu}}+\frac{\lambda}{\Lambda^{2\nu}}}= -\frac{\Lambda^{2\nu}}{\lambda}\frac{1}{\lambda(k^2/\Lambda^2)^\nu+1}
+\mbox{contact terms}.
\eeq{m53}

In K\"allen-Lehmann form, 
\beq
\widetilde{G}_{\nu,\lambda>0}(k^2)=+\frac{\Lambda^{2\nu}}{\pi}\int_0^\infty \frac{dm^2}{k^2+m^2}\frac{(m^2/\Lambda^2)^\nu (-\sin{\pi \nu})}{1+\lambda^2 (m^2/\Lambda^2)^{2\nu}+2\lambda (m^2/\Lambda^2)^\nu \cos{\pi\nu}}.
\eeq{m54}
Hence, the spectral density is positive-finite and the spectrum is free of ghosts and tachyons, thus providing a flow between the IR and UV fixed points. One can check that the IR fixed point of this flow is a CFT with an operator with $\Delta=d/2+|\nu|$, while at the UV fixed point $\Delta=d/2-|\nu|$. This is expected because the double-trace deformation is UV-irrelevant.

\subsubsection{$\lambda<0$}

Now for $\lambda<0$, 
\bea
\widetilde{G}_{\Delta,\lambda<0}(k^2)&=& \frac{1}{\frac{1}{k^{2\nu}}-\frac{|\lambda|}{\Lambda^{2\nu}}}\nonumber\\
&=& -\frac{\Lambda^{2\nu}}{|\lambda|}\frac{1}{|\lambda|(k^2/\Lambda^2)^\nu-1}+\mbox{contact terms}\nonumber\\
&=&+\frac{\Lambda^{2\nu}}{|\lambda|}\frac{1}{|\nu| |\lambda|^{1/\nu}\;(k^2/\Lambda^2-|\lambda|^{-1/\nu})}\nonumber\\
&\quad&+\frac{1}{\pi}\int_0^\infty \frac{dm^2}{k^2+m^2}\frac{m^{2\nu} (-\sin{\pi \nu})}{1+\lambda^2 (m^2/\Lambda^2)^{2\nu}-2|\lambda| (m^2/\Lambda^2)^\nu \cos{\pi\nu}}.
\eea{m55}
The two-point function has a pole of positive residue at 
$k^2/\Lambda^2=|\lambda|^{-1/\nu}$ on the positive real axis, signaling a 
(non-ghost) tachyon mode, while the smooth part
of the spectral density is positive definite.

\subsection{Case 3: $\nu>1$}
The case $\nu>1\not\in \mathbb{Z}^+$ has been studied in \cite{Andrade:2011dg} 
and \cite{Andrade:2011aa}. Those papers consider a UV completion of the
double-trace perturbation obtained by coupling a massive scalar to $O$. 
In our analysis we do not introduce any such scalar or any other {\em ad hoc}
UV completion. We use instead a K\"allen-Lehmann representation, which
can be obtained from those used in the previous subsection by replacing 
$(k^2/\Lambda^2)^\nu$ with $\pm(k^2/\Lambda^2)^\nu$, depending on the value of $\nu$.

For $\nu\in (2m+1,2m+2)$, $m\in\mathbb{Z}^+$ with deformation $\frac{\lambda}{2\Lambda^{2\nu}}O^2$, 
\beq
\widetilde{G}_{\nu,\lambda}(k^2)=\frac{1}{+\frac{1}{k^{2\nu}}+\frac{\lambda}{\Lambda^{2\nu}}}= -\frac{\Lambda^{2\nu}}{\lambda}\frac{1}{\lambda (k^2/\Lambda^2)^\nu+1}
+\mbox{contact terms}.
\eeq{m56}
The continuous part of the spectral density is positive-definite 
for both $\lambda>0$ and $\lambda<0$.

For $\lambda>0$, the simple poles appear at complex values of $k^2/\Lambda^2$ in conjugate pairs. The corresponding residues also form complex conjugate pairs, signaling tachyonic ghost modes. 

For $\lambda<0$, there is exactly one simple pole at real positive $k^2/\Lambda^2$, with negative residue, i.e. tachyonic ghost. 

These results are in agreement with the results of~\cite{Andrade:2011aa} for 
$2>\nu>1$.

On the other hand, for $\nu\in (2m,2m+1)$, $m\in\mathbb{Z}^+$,
\beq
\widetilde{G}_{\nu,\lambda}(k^2)=\frac{1}{-\frac{1}{k^{2\nu}}+\frac{\lambda}{\Lambda^{2\nu}}}= +\frac{\Lambda^{2\nu}}{\lambda}\frac{1}{\lambda (k^2/\Lambda^2)^\nu-1}
+\mbox{contact terms}.
\eeq{m57}
The spectral density is again positive-definite for any $\lambda$.

For $\lambda>0$, there is exactly one simple pole at real positive $k^2/\Lambda^2$, with positive residue, i.e. non-ghost tachyon. 

For $\lambda<0$, the simple poles are again at complex values of 
$k^2/\Lambda^2$ and appear in conjugate pairs with, 
conjugate residues. 
At least one pair has a negative real part, signaling a tachyonic ghost modes. 

Therefore, the deformed theory is not physical, as expected because at the 
putative UV fixed point the operator $O$ would have dimension 
$\Delta_-=d/2-\nu$, which is outside the unitary range.

\subsection{Case 4: $\nu=0$}
Next we consider the case $\Delta=d/2$. The two-point function is $\widetilde{G}_{\nu=0}(k^2)=-\ln{(k^2/\mu^2)}$. The appearance of the scale 
$\mu$ does not break conformal invariance, since rescaling $k$ amounts only to
changing the free energy $w[J]$ by a contact term $\sim J\cdot J$.

Introducing a (marginal) double-trace deformation $U[O]=+\frac{\lambda}{2} O^2$, we have 
\beq
\widetilde{G}_{\nu=0,\lambda}(k^2)=\frac{1}{-\frac{1}{\ln{(k^2/\mu^2)}}+\lambda}.
\eeq{m21}
The renormalization scale $\mu$ can be removed, as in~\cite{Witten:2001ua,Elitzur:2005kz}, by making the coupling constant $\lambda$ run with $\mu$. 
A convenient renormalization condition on $\lambda$ is to require that it 
diverges at some fixed scale $\Lambda$. This defines an RG flow of $\lambda$
\beq
\lambda\rightarrow \lambda(\mu)=+\frac{1}{\ln{(\Lambda^2/\mu^2)}}
\eeq{m22}
and so
\bea
\widetilde{G}_{\nu=0,\lambda}(k^2)&=&\frac{1}{-\frac{1}{\ln{(k^2/\mu^2)}}+\frac{1}{\ln{(\Lambda^2/\mu^2)}}}\nonumber\\
&=&\ln{(\Lambda^2/\mu^2)}-\ln^2{(\Lambda^2/\mu^2)}\frac{1}{\ln{(\Lambda^2/k^2)}}.
\eea{m23}
By performing the wave function renormalization  $O=ZO_R$, 
$Z=\ln(\Lambda^2/\mu^2)$, we obtain a $\mu$-independent renormalized two-point function
\beq
\widetilde{G}_{\nu=0,\lambda}^{\;(ren)}(k^2)=-\frac{1}{\ln{(\Lambda^2/k^2)}}.
\eeq{m24}
There is a simple pole at $k^2=\Lambda^2$ with positive residue, 
i.e. a tachyon mode. This is expected because this theory has a Landau pole for $\lambda>0$
at $\Lambda$ under our renormalization condition. For $\mu>\Lambda$, the 
coupling constant $\lambda(\mu)$ is negative and  asymptotically free, while $\lambda$ is 
positive and the theory is IR free for $\mu<\Lambda$. All of this is of course
in agreement with well known results for the $\lambda \phi^4$ theory in four dimensions.

\subsection{Case 5: $\nu=1$}
Now consider the special case, $\Delta=d/2+1$. 

Recall that
\beq
\widetilde{G}_{\nu=1}(k^2)=+k^2 \ln{(k^2/\mu^2)}.
\eeq{m25}
The two-point function of the alternative quantization $\Delta=d/2-1$ obtained
by a naive Legendre transformation (see Appendix) is
\beq
\widetilde{G}^{\:(?)}_{\nu=-1}(k^2)=-\frac{1}{k^2 \ln{(k^2/\mu^2)}}.
\eeq{m26}
Notice that the value of $\mu$ in equation~(\ref{m25}) does not spoil scale
invariance, since it can be changed by adding a contact terms. Instead
eq.~(\ref{m26}) is not scale invariant, since $\mu$ cannot be removed by local
counterterms.   
Moreover, the two-point function of an  operator saturating the unitarity 
bound is 
\beq
\widetilde{G}_{\nu=-1}(k^2)=\frac{1}{k^2}.
\eeq{m27}
So, eq.~(\ref{m26}) is not the two point function of a $\Delta=d/2-1$ conformal
field. In fact it is altogether unphysical, 
because it decays faster than $1/k^2$ at large $k^2$. 
The origin of this unphysical feature can be seen by representing the
two-point function in K\"allen-Lehmann form, because such representation
makes it manifest that there exists a simple pole at $k^2=\mu^2$ 
with negative residue:
\beq
\widetilde{G}^{\;(?)}_{\nu=-1}(k^2)=-\frac{1}{k^2-\mu^2}+\mathcal{O}((k^2-\mu^2)^0).
\eeq{m28}
In other words, the spectrum contains a tachyonic ghost mode. Notice that $\mu$ is a physical scale, not an auxiliary one that can be removed by local counterterms. In fact $\mu$ is physical even at the UV fixed point, as 
pointed out earlier.

Using Cauchy's formula one obtains
\beq
-\frac{1}{k^2 \ln{(k^2/\mu^2)}}=-\frac{1}{k^2-\mu^2}+\int_0^\infty dm^2 \frac{1}{k^2+m^2}\frac{1}{(m^2/\mu^2)(\ln^2{(m^2/\mu^2)}+\pi^2)}
\eeq{m29}
with
\beq
\int_0^\infty\frac{dm^2}{m^2(\ln^2{m^2}+\pi^2)}=1.
\eeq{m30}

Now add the double-trace deformation $U[O]=\frac{\lambda}{2\Lambda^2}O^2$. 

For $\lambda>0$, 
\bea
&\quad&\widetilde{G}_{\nu=1,\lambda>0}(k^2)\nonumber\\
&=&\frac{1}{\frac{1}{k^2 \ln{(k^2/\mu^2)}} +\frac{\lambda}{\Lambda^2}} \nonumber\\
&=& -\frac{\Lambda^2}{\lambda}\frac{1}{1+\lambda \frac{k^2}{\Lambda^2} \ln{(k^2/\mu^2)}}+\mbox{contact terms} \nonumber\\
&=&\textsc{two simple poles}+\int_0^\infty dm^2\frac{1}{k^2+m^2}\frac{m^2}{\pi^2+\left(1-(\frac{\lambda}{\Lambda^2})m^2\ln{(m^2/\mu^2)}\right)^2}.
\eea{m33}
The branch-cut in the complex $k^2/\Lambda^2$ plane is logarithmic. The spectral density is positive-definite, but now there are two simple poles.

For $\lambda\mu^2/\Lambda^2>e$, the poles are at real and positive $k^2$. The pole at larger $k^2$ has a negative residue and that at the smaller $k^2$ a positive residue, with the latter pole approaching 0 as $\lambda \mu^2/\Lambda^2\rightarrow \infty$.

For $\lambda\mu^2/\Lambda^2<e$, the poles are complex and are conjugates of each other. The residues have a positive real part and complex conjugate imaginary parts.

For $\lambda<0$, 
\bea
\widetilde{G}_{\nu=1,\lambda<0}(k^2)&=&\frac{1}{\frac{1}{k^2 \ln{(k^2/\mu^2)}} - \frac{|\lambda|}{\Lambda^2}} \nonumber\\
&=&-\frac{\Lambda^2}{|\lambda|}\frac{-|\lambda| \frac{k^2}{\Lambda^2} \ln{(k^2/\mu^2)}}{1-|\lambda| \frac{k^2}{\Lambda^2} \ln{(k^2/\mu^2)}}\nonumber\\
&=& +\frac{\Lambda^2}{|\lambda|}\frac{1}{1-|\lambda| \frac{k^2}{\Lambda^2} 
\ln{(k^2/\mu^2)}}+\mbox{contact terms},
\eea{m31}
which again decays faster than $1/k^2$ at large $k^2$ for any $|\lambda|/\Lambda^2 \neq 0$, so it is again unphysical.

A contour integration gives 
\beq
\widetilde{G}_{\Delta=d/2+1,\lambda<0}(k^2)=\textsc{one simple pole}+\int_0^\infty dm^2 \frac{1}{k^2+m^2}\frac{m^2}{\pi^2+\left(1+(\frac{|\lambda|}{\Lambda^2})\;m^2\;\ln{(m^2/\mu^2)}\right)^2}
\eeq{m32}
in which the spectral density in the second term is positive. The pole is at real and positive $k^2$ with negative residue, signaling the propagation of a ghost tachyon mode.

Therefore, one concludes that the flow to the theory with $\Delta=d/2-1$ in the UV is unphysical for all values of $\lambda$. Moreover, the very
fact that the two-point function in eq.~(\ref{m26}) is non-unitary shows that
the UV limit $\Lambda\rightarrow 0$ is meaningless in this case.

The singleton point is reached the limit $\lambda\rightarrow 0$, 
$\mu \exp(-1/\lambda^2)=\mbox{constant}$. This limit does decouple all ghost and physical states and leads to a two-point 
function $\propto 1/k^2$, but it cannot be achieved as an RG trajectory. A different singular limit leading to the singleton is
described in~\cite{Ohl:2012bk}.

\section{Summary}
After reviewing the method that allows to find the two-point function of certain primary operators $O$ in large-$N$ theories 
deformed by interactions proportional to $O^2$, we studied the RG flow that is determined by the deformation. To get an
understanding that goes beyond what is available in the (vast) existing literature on the subject, we used the 
K\"allen-Lehmann representation of the two-point function. This representation allows for a clear and unambiguous detection of unphysical features such as tachyons or ghosts. The
presence of such features has a natural interpretation in the case $\nu<-1$, 
where one of the possible dimensions
for the operator $O$, $\Delta_-=d/2 -\nu$, violates the unitarity bound $\Delta\geq d/2-1$. 

We also found that the K\"allen-Lehmann representation automatically  contains extra massive scalars, signaled by simple 
scalar poles in the two-point function. We can say that the K\"allen-Lehmann representation ``integrates in'' massive 
scalar. One result of our study is that, when the 
perturbation $O^2$ is relevant, the extra massive
scalar found using the K\"allen-Lehmann representation is physical 
when $\lambda>0$ and tachyonic when $\lambda<0$. The example of the bosonic 
$O(N)$ vector model in three dimensions shows that this is the 
expected behavior, because the potential 
$(\lambda/2N)(\sum_{a=1}^N\phi^a\phi^a)^2$ 
is stable only for $\lambda>0$. On the other hand, the massive scalar
used in~\cite{Andrade:2011dg,Andrade:2011aa} to define a UV completion of 
double-trace deformations has the opposite behavior: it is 
tachyonic for $\lambda>0$ and physical for $\lambda<0$. One possible reason 
for the disagreement is that the UV completion used 
in~\cite{Andrade:2011dg,Andrade:2011aa} can be pathological in
the IR. This is manifest in the case of the $O(N)$ vector model, where the 
scalar potential is unbounded below for either sign of $\lambda$.

An especially interesting case is $\nu=1$, because 
the conformal dimensions allowed by the alternative quantization of
ref.~\cite{Klebanov:1999tb} appears to saturate the unitarity bound $\Delta=d/2-1$. 
This would be a free field that does not have a dual in a putative 
semiclassical AdS gravity in $d+1$ dimensions. 
The flow generated by a double trace perturbation would define a
theory where $\Delta=d/2+1$ flows to the problematic UV fixed point with 
$\Delta=d/2-1$. We found that the flow is
unphysical, because the two-point function of the operator $O$ always
contains unphysical states and the UV fixed point itself is unphysical. 

Finally we should remark that our analysis agrees with ref.~\cite{Bashmakov:2016pcg}, which studies 
double-trace deformations involving two {\em different} operators. Here we will restrict our analysis to the most interesting case
that one of the two operators, $O_1$,  saturates the unitarity bound ($\Delta_1=d/2-1$) while the other, $O_2$, has 
dimension $\Delta_2>\Delta_1$, $\Delta_2<d/2$. Ref.~\cite{Bashmakov:2016pcg} studies a relevant flow from 
the UV, where $O_1$ saturates the unitarity bound, to the IR. It is thus quite different from the situation considered in this 
paper, which considers an irrelevant flow to a putative UV fixed point. Nevertheless, the flow can be studied easily using the methods described in this paper. 
The deformation studied in~\cite{Bashmakov:2016pcg} is 
\beq
N\int d^dx fO_1 O_2 +g_1 O_1^2 + g_2O^2_2, \qquad f,g_1,g_2\in \mathbb{R}. 
\eeq{add0}
One can check that when $g_1g_2>f^2, g_1>0,g_2>0$ the flow is physical\footnote{We thank M. Bertolini 
for pointing this out to us.};
on the other hand, whenever $g_1\neq 0$ one is simply giving a mass to a free scalar, so the flow is rather trivial: 
a massive scalar
decouples in the IR. So, let us consider the case $g_1=0$. 

At large $N$ the deformation changes
the two point functions $\langle \tilde{O}_i(k) O_j(0)\rangle$, $i,j=1,2$ as
\beq
\langle \tilde{O}_i(k) O_j(0)\rangle=\begin{pmatrix} k^2 & f \\ f & k^{d-2\Delta_2} +g_2\end{pmatrix}^{-1}=
{1\over k^{d+2-2\Delta_2} +g_2k^2 -f^2}\begin{pmatrix} k^{d-2\Delta_2} +g_2 & -f \\ -f & k^2 \end{pmatrix}.
\eeq{add1}
In the extreme infrared, $k^{2(1+d/2-\Delta_2)}\ll f^2$  
\beq
\langle \tilde{O}_i(k) O_j(0)\rangle=-{k^{d-2\Delta_2}\over f^2} \begin{pmatrix} 1 & -{k^2\over f} \\ -{k^2\over f }& {k^4\over f^2} \end{pmatrix}+\mbox{contact terms}.
\eeq{add2}
The matrix in~(\ref{add2}) has rank one and is independent of $g_2$, meaning that $O_1= f\Box O_2$ in the extreme infrared, where $O_2$ has
dimension $d-\Delta_2$. All this is in perfect agreement with ref.~\cite{Bashmakov:2016pcg}. 
We conclude by observing that, while some of the UV-complete theories with $g_1>0,g_2>0$ are physical, those
with $g_1=0$ are plagued by unphysical states, since the two point function in~(\ref{add1}) has, 
among other unpleasantnesses, a tachyonic pole for any value of $g_2$. The RG flow generated by the 
deformation~(\ref{add0}) with $g_1=0$ is therefore unphysical, according to our general definition.

\subsection*{Acknowledgements} 
We thank M. Bertolini for useful comments. C.Y.Y. and M.P. are supported in part by NSF grant PHY-1316452. 

\section*{Appendix: Alternative Quantization in AdS/CFT}
\setcounter{equation}{0}
\renewcommand{\theequation}{A.\arabic{equation}} 
In this appendix we review the standard $AdS_{d+1}/CFT_d$ correspondence and the relationship between the two 
choices of quantization explained in~\cite{Klebanov:1999tb}. 
Euclidean $AdS_{d+1}$ admits the Poincar\'e metric
\beq
ds^2=\frac{1}{z^2}\left(dz^2+\sum_{i=1}^ddx_i^2 \right),
\eeq{m34}
where we set the $AdS_{d+1}$ radius to one. A scalar field $\phi(z,\vec{x})$ of mass $m$ in the bulk has the asymptotic form near the boundary ($z\ll 1$)
\beq
\phi(z,\vec{x})=z^{d-\Delta}(\phi_0 (\vec{x})+O(z^2))+z^\Delta \left(\frac{A(\vec{x})}{2\Delta-d}+O(z^2)\right)
\eeq{m35}
in which $\Delta=\Delta_\pm=\frac{d}{2}\pm\sqrt{\frac{d^2}{4}+m^2}\equiv \frac{d}{2}\pm\nu$ for $\Delta \neq d/2$, $\nu \notin \mathbb{Z}$.
 
Evaluating the on-shell action and discarding possible contact terms, one finds for $\Delta=\Delta_+>d/2$
\bea
S_{AdS_{d+1}}[\phi_0]&=&-\frac{1}{2}\int d^dx\; \phi_0(x) A(x)\nonumber\\
&=&-\frac{1}{2}(2\Delta-d)\pi^{-d/2}\frac{\Gamma(\Delta)}{\Gamma(\Delta-d/2)} \int d^dx \int d^dx' \frac{\phi_0(x)\phi_0(x')}{|x-x'|^{2\Delta}}.
\eea{m36}
 
In the case $\Delta=d/2$,
\beq
\phi(z,\vec{x})= z^{d/2}(\ln{(z/z_0)}\phi_0(\vec{x})+A(\vec{x})+O(z^2)).
\eeq{m37}
This asymptotic behavior implies that \cite{Freedman:1998tz}
\beq
S_{AdS_{d+1}}[\phi_0]=-\frac{1}{2} \frac{\Gamma(d/2)}{2\pi^{d/2}}\int d^dx \int d^dx' \; \phi_0(\vec{x}) \frac{1}{|\vec{x}-\vec{x'}|^{d}}\phi_0(\vec{x}).
\eeq{m38}

The  AdS/CFT correspondence then reads
\beq
\exp{(-N^* S_{AdS_{d+1}}[\phi_0])}=\left\langle \exp{\left(-N^* \int \phi_0\; {{}}{O}\right)} \right\rangle_{CFT_d}=\int [dO]\; e^{-I_{CFT_d}[{O}]-\int N^* \phi_0 {{}}{O}}.
\eeq{m39}
We have multiplied the exponent by a factor of $N^* \gg 1$ 
such that {$\phi_0$ and $O$ are both $\mathcal{O}(1)$. For the O(N) model, $N^*=N$. }

The corresponding operator ${{}}{O}(x)$ is of conformal dimension $\Delta=\Delta_+$. Unitarity requires $\Delta\geq d/2-1 $ \cite{Breitenlohner:1982jf}.

By virtue of eqs.~(\ref{m36}, \ref{m38}), we identify $-N^* S_{AdS_{d+1}}[\phi_0]$
with $W[J]$, the generating functional of connected correlators of the 
boundary CFT defined in the
main body of the paper. The boundary field $\phi_0(x)$ is identified with 
the source $J$ of the boundary conformal operator ${{}}{O}(x)$ as
\beq
J=-\phi_0,
\eeq{m40}
and 
\beq
{O} =+\frac{\delta S_{AdS_{d+1}}[\phi_0]}{\delta \phi_0} =-A.
\eeq{m41}

The two-point function of ${{}}{O}(x)$ can be immediately read off from the action. For $\Delta>d/2$
\beq
N^*\langle O(x)O(0) \rangle=(2\Delta-d)\pi^{-d/2}\frac{\Gamma(\Delta)}{\Gamma(\Delta-d/2)}\frac{1}{|x|^{2\Delta}}
\equiv G_\nu (|x|)
\eeq{m42}
and  for $\Delta=d/2$, $N^*\langle {O}(x){{}}{O}(0) \rangle=\frac{\Gamma(d/2)}{2\pi^{d/2}}\frac{1}{|x|^{2\Delta}}$.

The two-point function in momentum-space representation in the case $\Delta=d/2+1$ contains gamma functions in the prefactor that appear divergent, see 
eq.~(\ref{m6}). However, one notes that
\bea
&\;&\left( \frac{d}{dk^2} \right)^2 N^* \langle \widetilde{{{}}{O}}(k){{{}}{O}}(0) \rangle \nonumber\\
&=&\left( \frac{d}{dk^2} \right)^2 \left[\left( (2\Delta-d)\pi^{-d/2}\frac{\Gamma(\Delta)}{\Gamma(\Delta-d/2)} \right)\int d^dx\; e^{-ik\cdot x} \frac{1}{x^{2\Delta}}\right]_{\Delta=d/2+1}\nonumber\\
&=& 2\pi^{-d/2}\Gamma(d/2+1) \int d^dx\; e^{-ik\cdot x} \frac{1}{x^{2((d/2+1)-2)}}\nonumber\\
&=& 2\pi^{-d/2}\Gamma(d/2+1) \int d^dx \;e^{-ik\cdot x} \int_0^\infty \frac{dt}{t} t^{d/2-1} e^{-x^2 t}/\Gamma(d/2-1)\nonumber\\
&=& 2\pi^{-d/2}\Gamma(d/2+1) \int_0^\infty \frac{dt}{t}\;t^{d/2-1}\left( \frac{\pi}{t} \right)^{d/2}\;e^{-\frac{k^2}{4t}}/\Gamma(d/2-1)\nonumber\\
&\overset{s=\frac{1}{t}}{=}& 2\pi^{-d/2}\Gamma(d/2+1)\; \frac{\pi^{d/2}\int_0^\infty \frac{ds}{s}se^{-\frac{k^2}{4}s}}{\Gamma(d/2-1)}\nonumber\\
&=&2d(d-2)\frac{1}{k^2}.
\eea{m43}
Integrating twice, getting rid of contact terms, and introducing
a fictitious scale $\mu$, one arrives at
\beq
\widetilde{G}_{\nu=1}(k^2)=N^* \langle \widetilde{O}(k) O(0) \rangle=2d(d-2) k^2 \ln{(k^2/\mu^2)},
\eeq{m44}
with a positive prefactor for $d>2$. 

For $d/2+1\geq \Delta > d/2$, the alternative quantization in which $\Delta=\Delta_-$ is also allowed by unitarity. To obtain a correspondence to another CFT 
where the operator $O$ has conformal dimension $\Delta_-$, 
one needs to exchange the roles of $\phi_0$ and $A$. 
Since $\phi_0$ and $A$ are conjugate variables, the exchange is done by a 
Legendre transformation~\cite{Klebanov:1999tb}.

Define the effective ``intensive'' action
\beq
\gamma[O]=S_{AdS_{d+1}}[\phi_0]-\phi_0 O
\eeq{m45}
such that 
\beq
\frac{\delta(\gamma[O])}{\delta O}=-\phi_0
\eeq{m46}

Recall that
\beq
-\widetilde{A}(k)= \widetilde{O}=+\frac{\delta S_{AdS_{d+1}}[\widetilde{\phi_0}]}{\delta\widetilde{\phi_0}}=-\widetilde{G}_\nu (k^2)\widetilde{\phi_0}(-k),
\eeq{m48}
the effective action of the (undeformed) CFT is
\bea
\gamma_0[\widetilde{A}]&=&+\frac{1}{2}\int \frac{d^dk}{(2\pi)^d}\;\widetilde{A}(k)\frac{1}{\widetilde{G}_{\nu}(k^2)}\widetilde{A}(-k)\nonumber\\
&=& -\frac{1}{2}\int \frac{d^dk}{(2\pi)^d}\;\widetilde{A}(k)\widetilde{G}_{-\nu}(k^2)\widetilde{A}(-k) .
\eea{m49}
Instead of interpreting $\gamma_0[\tilde{A}]$ as the effective action for ${{}}{O}$, one can interpret it as the free energy of the CFT operator ${{}}{O}'$ of conformal dimension $\Delta_-$ because the two-point function in the 
alternative quantization is $\widetilde{G}_{-\nu}(k^2)=-1/\widetilde{G}_{\nu}(k^2)$.

%%%%%%%%%%%%%%%%%%%%%%%%%%%%%%%%%%%%%%%%%%%%%%%%%%%

\end{document}